\documentclass[9pt,english,preview,sort&compress, letterpaper]{elsarticle}
\usepackage{graphicx}
\usepackage{amssymb,latexsym,amsthm,amsmath}
\usepackage{epstopdf}
\usepackage[applemac]{inputenc}
\usepackage{palatino, url, multicol}
\usepackage[mathscr]{eucal}
\usepackage[width=7cm,font=small]{caption}
\usepackage{rotating}
\usepackage{fullpage}
\usepackage{array}
\usepackage[usenames,dvipsnames]{color}
\usepackage{xfrac}
\biboptions{super,comma,square,compress}
\title{Algorithm to compute the electric field gradient tensor in ionic crystals}

\cortext[cor1]{Corresponding author: jorge\_hdz@ciencias.unam.mx}
\cortext[cor2]{Principal corresponding author: rgomez@servidor.unam.mx, tel: (+52) (55) 5622-4849}

\address[dir]{Facultad de Ciencias, Universidad Nacional Autónoma de México\\
Circuito Exterior, Ciudad Universitaria, D.F., México, 04510}

\author[dir]{J.J. Hernández-Gómez\corref{cor1}}  
\ead{jorge_hdz@ciencias.unam.mx}

\author[dir]{V. Marquina} 
\ead{marquina@servidor.unam.mx}

\author[dir]{R.W. Gómez\corref{cor2}} 
\ead{rgomez@servidor.unam.mx}

\begin{document}

\begin{abstract}
A simple algorithm and a computational program to numerically compute the electric field gradient and the concomitant quadrupolar nuclear splitting is developed for an arbitrary ionic crystal. The calculations are performed using a point charge model. The program provides three different ways for the data input: by Bravais lattices, by lattice parameters, or by introducing any spatial structure. The program calculates the components of the electric field gradient, the asymmetry parameter and the quadrupolar splitting for a given number of nearest neighbors with respect to the nuclear charge as origin. In addition, the program allows the use of different Sternheimer antishielding factors.
\end{abstract}


\maketitle

\section{Introduction}
\noindent The electrostatic energy $W$ due to the interaction of a nuclear charge distribution $\rho(\b{r})$ and the electrostatic potential $V(\b{r})$ generated by its electric environment is given by 
\begin{equation}
W= \int_{vol} \rho(\b{r}) V(\b{r}) d^3\tau \label{E0}\\  
\end{equation}
where $d^3\tau$ is the volume element and $\b{r}=(x_{1},x_2,x_3)$ are spatial coordinates. The integral is calculated over the nucleus volume. A suitable way to evaluate it is to make a multipole expansion of the electrostatic potential $V(\b{r})$ around the center of charge of the nucleus as origin, assuming that $V(\b{r})$ is a slowly varying function over the nuclear region. Expanding in a Taylor serie around the nucleus center of charge, one obtains\textcolor{NavyBlue}{\cite{Blatt, Dunlop}}
\begin{equation}
V(\b{r})= V(\b{0}) + \b{r}\cdot (\nabla V)_{\b{r}=\b{0}} + \frac{1}{2} \sum_i\sum_j x_i x_j \left( \frac{\partial^2V}{\partial x_i \partial x_j} \right) + \cdot\cdot\cdot \label{E1}\\  
\end{equation}

The relevant terms in this expansion are the first and third terms, due to the fact that the second one is zero\footnote{As well as all the odd terms in the expansion.} because, when multiplied by the nuclear charge, it represents the interaction of the nuclear dipole moment (which is zero) with the external electric field, $\vec{E}$. The next non-zero terms are several orders of magnitude smaller than the third one\textcolor{NavyBlue}{\cite{Jackson}} so, in a good approximation, the interaction energy can be expressed as
\begin{equation}
W= \frac{1}{2}\sum_{j,k=1}^{3}V_{jk}Q_{jk} \label{E1}\\  
\end{equation}
where $V_{jk}$ are the electric field gradient (EFG) tensor components, and $Q_{ij}$ are the quadrupolar nuclear moment components; both are second rank tensors. Choosing a principal axis system for the EFG tensor, the interaction energy can be expressed as the sum of two terms

\begin{equation}
W= \frac{1}{6}\sum_{j=1}^{3}V_{jj}\int_{vol}\rho_{N}(\b{r})r^{2}d^{3}\tau + \frac{1}{2}\sum_{j=1}^{3}V_{jj}\int_{vol}\rho_{N}(\b{r})(x_{j}^{2}-\frac{r^{2}}{3})d^{3}\tau \label{E2}  
\end{equation}

The first term, called isomer shift, represents the effect due to the nucleus size. The second one corresponds to the so called quadrupolar nuclear splitting $\Delta Q$, so the interaction hamiltonian between the nuclear quadrupolar moment $\tilde{Q}$ and the EFG tensor $\nabla\vec{E}$, with respect to an arbitrary axes system with origin in the nuclear charge centroid, is given by
\begin{equation}
\widehat{\mathscr{H}}=-\frac{1}{6}e\tilde{Q}\otimes\nabla\vec{E} \label{E3}\\  
\end{equation}
where $\otimes$ denotes tensorial product. Considering the $z$ axis along the largest component of the EFG ($V_{zz}=eq$) and the Laplace equation, the hamiltonian (\ref{E3}) is transformed through the Wigner-Eckart theorem\textcolor{NavyBlue}{\cite{Messiah}} into
\begin{equation}
\widehat{\mathscr{H}}=\frac{e^{2}qQ}{4I(2 I -1)}[3I_{z}^{2}-I^{2}+\eta(I_{x}^{2}-{I}_{y}^{2})] \label{E4},  \\
\end{equation}
where ${I}^{2}$, ${I}_{x}$, ${I}_{y}$ and ${I}_{z}$ are the nuclear spin magnitudes, and

\begin{equation}
\eta=\frac{V_{xx}-V_{yy}}{V_{zz}} \label{E5}\\
\end{equation}
is the so called asymmetry parameter, which indicates how much the electric potential departs from spherical symmetry. An analytical solution of (\ref{E4}) can only be obtained for the $I=\sfrac{3}{2}$ case\textcolor{NavyBlue}{\cite{Greenwood}}. By far, the most used isotope in Mössbauer spectroscopy is $^{57}$Fe, for which the useful transition is $I=\sfrac{3}{2} \to I=\sfrac{1}{2}$ and, in what follows, we will restrict to this case. The analytical solution is:
\begin{equation}
E=\frac{e^{2}qQ}{4I(2 I-1)}[3 I_{z}^{2}-I^2]\sqrt{1-\frac{\eta^{2}}{3}} \label{E6}\\
\end{equation}

That is, the nuclear $I = \sfrac{3}{2}$ energy level is split into two levels $(\pm\sfrac{3}{2}$ and $\pm\sfrac{1}{2})$ and the ground level $I = \sfrac{1}{2}$ stays degenerated. This gives rise to two absorption lines in the Mössbauer spectrum separated by an energy
\begin{equation}
\Delta Q=\frac{eV_{zz}Q}{2}\sqrt{1+\frac{\eta^{2}}{3}} \label{E7}\\
\end{equation}
which is called the quadrupolar nuclear splitting.

\subsection{Electric Gradient Tensor} 

\noindent In rectangular coordinates, the EFG of a set of $n$ point charges is:
\begin{equation}
V_{x_{i}x_{j}}=\sum_{k=1}^{n}q_{k}\left(\frac{r_{k}^{2}\delta_{ij}-3x_{i_{k}}x_{j_{k}}}{r_{k}^{5}}\right) \label{E8} \\
\end{equation}
where $q_{k}$ and $\b{r}_{k}=(x_{1_{k}},x_{2_{k}},x_{3_{k}})$ are the respective charge and position of the $k^{th}$ ion. The electric interaction of the nucleus with its surroundings has two different origins: the charge density of the electrons of the nucleus under study, and the ligands of the crystal lattice\textcolor{NavyBlue}{\cite{Fraunfelder, Greenwood}}.

For ionic crystals, the main contribution to the EFG comes from those ions directly coordinated to the nucleus. Considering that the interatomic distances in a crystal are pretty much larger than the displacements due to the ion vibrations, a good approximation to the EFG can be done using equation (\ref{E8}) with a point charge model. The contribution to the EFG due to the electronic distribution requires a complex calculation that involves not only the knowledge of the electronic wave functions of the atom, but also shielding and antishielding effects and polarization of the electronic distribution due to other charges near by. This type of calculations retreat from the scope of this work, so their effect will be taken into account trough two parameters $R$ and $\gamma_\infty$, known as the  shielding and antishielding Sternheimer factors\textcolor{NavyBlue}{\cite{Sternheimer1, Sternheimer2, Sternheimer3, Sternheimer4, Sternheimer5}}, to obtain 
\begin{equation}
q=q_{lig}(1-\gamma_{\infty})+q_{val}(1-R)\label{E9}\\
\end{equation}

The first term in equation (\ref{E9}) corresponds to the ligand contribution, while the second one constitutes the valence contribution. However, in ionic crystals the last one can be neglected. The following program is developed for this case.

\section{Structure of the computational program}

\noindent The program was focused as an useful tool in a Mössbauer spectroscopy laboratory, so it computes the components of the EFG tensor and the quadrupolar splitting for a $^{57}$Fe nucleus by default. However, it is able to work with any nucleus, just by introducing its respective quadrupolar moment value.

In order to compute the EFG in a great number of crystalline lattices, three different input data modes were developed, allowing for a wide range of applications. Those modes are briefly described here:

\begin{itemize}
\item \emph{N arbitrary ions}: In this section of the program, the coordinates and valences of each ligand constituting the crystalline array are inputted manually. The spatial distribution of the ions can be totally arbitrary. The algorithm can handle a number of ions as large as necessary, being this number, of course, finite.

\item \emph{Bravais lattices:} Here, an election of one of the fourteen possible Bravais Lattices in three dimensions is made, just by introducing the parameter(s) that define such lattice. The program allows to select the place in the lattice in which the EFG will be computed.

\item \emph{Lattice parameters:} When available, one can input the values of the lattice parameters, so the program identifies the respective lattice, reconstructing it in order to carry out the computations.
\end{itemize}

The program was developed in a structured computational language, so there is a main module calling different functions and subroutines.

\subsection{Functions} There are three functions defined in the program.

\subsubsection{$R$} The function $R$ calculates the euclidean rectangular distance between the $i^{th}$ ligand coordinates and the nucleus under study taken as the origin.

\subsubsection{V} This function calculates the $V_{x_ix_i}$ component of the EFG tensor in principal axes for the $i^{th}$ ion with equation (\ref{E8}), for each value of $k$, and where $q_k$ is the valence charge of the ligand, $q_{lig}$.

\subsubsection{DQ} This function computes the value of the quadrupolar splitting as a function of $V_{zz}$ and the asymmetry parameter, through equations (\ref{E7}) and (\ref{E9}), leaving the result in terms of the $(1-\gamma_\infty)$ factor, without considering the valence contribution in the total charge.


\subsection{Main module} Here, the physical constants to be used in the program are defined. It establishes the value of the quadrupolar moment $Q$ to be used\footnote{The program uses the value $Q=16$b of the quadrupolar nuclear moment for the $^{57}$Fe, recently reported by Martínez-Pinedo et al\textcolor{NavyBlue}{\cite{Martinez-Pinedo}}.}, and the ionic configuration to be worked out, which are:

\subsubsection{N arbitrary ions}

\begin{itemize}
\item Introduce the number $N$ of ligands to be considered in the computation.
\item Introduce the three coordinates (in angstroms) and the valence of each ion.
\item The distance to the origin is computed for each ion, via the $R$ function.
\item The components of the EFG are calculated, adding\footnote{Due to the superposition principle of the electric potential.\label{Foot0}} to each one the contribution of each ion through function $V$.
\item The largest component of the EFG is assigned to $\vert V_{zz}\vert$, and $\vert V_{yy}\vert \ge \vert V_{xx}\vert$.
\item The asymmetry parameter is computed with equation (\ref{E5}).
\item The value of the quadrupolar splitting is computed with the function $DQ$.
\item The results of the EFG components, the asymmetry parameter and the quadrupolar splitting are shown on screen and saved in a file.
\end{itemize}

\subsubsection{Bravais Lattices}

\begin{itemize}
\item Choose one of the seven possible groups in three dimensions.
\item Select the lattice to be taken into account\footnote{If possible, choose between the simple, the body centered, the two faces centered or the face centered correspondent structure.} and introduce the parameter(s) that define it. Then choose the number of nearest neighbors to be deemed, the valence of each layer of neighbors, and the position in the structure in which the EFG is to be computed (in the center or the vertex of the structure).
\item With this information, the program reckons the coordinates of the ligands in the lattice, through the algorithm presented in the next section. Once the coordinates are determined, the distance of each ligand to the origin is computed. In order to identify and count the layers of nearest neighbors, the information of all the generated neighbors is ordered and displayed in growing distances to the origin\footnote{The ordering process is carried out with a bubble ordering algorithm, which does not compromises the efficiency of the program because the lists of neighbors to be ordered are not generally too large in standard calculations in solid state.\label{Foot0}}. The components of the EFG are calculated for the chosen layers of neighbors.
\item The largest component of the EFG assigned to $\vert V_{zz}\vert$, and $\vert V_{yy}\vert \ge \vert V_{xx}\vert$.
\item The asymmetry parameter is computed with equation (\ref{E5}).
\item The value of the quadrupolar splitting is reckoned with the function $DQ$.
\item The results of the EFG components, the asymmetry parameter and the quadrupolar splitting are shown on screen and saved in a file.
\end{itemize}

\subsubsection{Lattice parameters}

\begin{itemize}
\item Introduce the six lattice parameters, ($a$, $b$, $c$) and ($\alpha$, $\beta$, $\gamma$), in angstroms and degrees respectively.
\item The program identifies the lattice that corresponds with the lattice parameters introduced, and with the information of the lattice, the program proceeds in the same form than in the previous section, \emph{Bravais lattices}.
\end{itemize}

\section{Algorithm}

\noindent In what follows, the main algorithm used by the sections \emph{Bravais lattices} and \emph{lattice parameters} to find the points in the lattice where the ligands are to be considered for the computations, is described:

\begin{enumerate}
\item Select the number of nearest neighbors to be considered.
\item 
\begin{itemize}
\item If all the neighbors have the same valence, introduce it.
\item If not, introduce the valence of each layer of neighbors.
\end{itemize}
\item Choose to compute the EFG in the center or in the vertex of the structure.
\item With the six lattice parameters ($a$, $b$, $c$) y ($\alpha$, $\beta$, $\gamma$), the rectangular components of the crystallographic axes are calculated through the next transformation equations, obtained in the appendix \ref{ap:coords}

\begin{equation}
a_x=a \label{E10}\\
\end{equation}
\begin{equation}
b_x=b\cos\gamma \notag \\
\end{equation}
\begin{equation}
b_y=b\sin\gamma \notag \\
\end{equation}
\begin{equation}
c_x=c\cos\beta \notag \\
\end{equation}
\begin{equation}
c_y=c(\cos\alpha\csc\gamma-\cos\beta\cot\gamma) \notag \\
\end{equation}
\begin{equation}
c_z=c\,\sin\beta\sqrt{1-(\cos\alpha\csc\beta\csc\gamma-\cot\beta\cot\gamma)^2} \notag \\
\end{equation}

\item If the studied nucleus is centered in the body, the coordinates of the ligands are calculated as follows:
\begin{itemize}
\item If the structure is simple (SC, ST, SO, SM, triclinic, trigonal or hexagonal)\footnote{These are common abbreviations in crystallography. SC: Simple Cubic; ST: Simple Tetragonal; SO: Simple Orthorhombic;  SM: Simple Monoclinic; BCC: Body Centered Cubic; BCT: Body Centered Tetragonal; BCO: Body Centered Orthorhombic; FCC: Face Centered Cubic; FCO: Face Centered Orthorhombic; 2FCO: Two Face Centered Orthorhombic; 2FCM: Two Face Centered Monoclinic. \label{Foot2}}, the coordinates of the $i^{th}$ ion are computed through equations (\ref{E11}), where the numbers $n_1$, $n_2$ and $n_3$ are whole numbers in the interval $[-m,m]$\footnote{The number $m$ (chosen in the step a)) depends on the number of nearest neighbors to be considered in the computation. For example, $m=2$ is enough to find the third nearest neighbors.\label{Foot1}}. 

\begin{equation}
x_i = (n_1+\frac{1}{2})a_x+(n_2+\frac{1}{2})b_x+(n_3+\frac{1}{2})c_x \label{E11} \\
\end{equation}
\begin{equation}
y_i = (n_2+\frac{1}{2})b_y+(n_3+\frac{1}{2})c_y \notag \\
\end{equation}
\begin{equation}
z_i = (n_3+\frac{1}{2})c_z \notag \\
\end{equation}
\item If the structure is body centered (BCC, BCT or BCO)$^{\ref{Foot2}}$, the coordinates of the $i^{th}$ ion are computed through equations (\ref{E12}), the same way as in simple structures, but excluding the point $(0,0,0)$. 
\begin{equation}
x_i = n_1a_x+n_2b_x+n_3c_x \label{E12} \\
\end{equation}
\begin{equation}
y_i = n_2b_y+n_3c_y \notag \\
\end{equation}
\begin{equation}
z_i = n_3c_z \notag \\
\end{equation}
\item If the structure is face centered (FCC or FCO)$^{\ref{Foot2}}$, the coordinates of the $i^{th}$ ion are computed through equations (\ref{E13}), (\ref{E14}) and (\ref{E15}) for the ligands in the faces parallel to the crystallographic planes, where the numbers $n_1$, $n_2$ y $n_3$ are whole numbers in the interval $[-m,m]$$^{\ref{Foot1}}$ and such that $n_1$ y $n_2$ can not be zero. 
\begin{equation}
x_i = \frac{1}{2}n_1a_x+\frac{1}{2}n_2b_x+n_3c_x \label{E13} \\
\end{equation}
\begin{equation}
y_i = \frac{1}{2}n_2b_y+n_3c_y \notag \\
\end{equation}
\begin{equation}
z_i = n_3c_z \notag\\
\end{equation}
\begin{equation}
x_i = \frac{1}{2}n_1a_x+\frac{1}{2}n_2c_x+n_3b_x \label{E14} \\
\end{equation}
\begin{equation}
y_i = \frac{1}{2}n_2c_y+n_3b_y \notag \\
\end{equation}
\begin{equation}
z_i = \frac{1}{2}n_2c_z \notag\\
\end{equation}
\begin{equation}
x_i = \frac{1}{2}n_1c_x+\frac{1}{2}n_2b_x+n_3a_x \label{E15} \\
\end{equation}
\begin{equation}
y_i = \frac{1}{2}n_2b_y+\frac{1}{2}n_1c_y \notag \\
\end{equation}
\begin{equation}
z_i = \frac{1}{2}n_1c_z  \notag\\
\end{equation}
\item If the structure is two face centered (2FCO or 2FCM)$^{\ref{Foot2}}$, the coordinates of the $i^{th}$ ion are computed through equations (\ref{E16}), where the numbers $n_1$, $n_2$ y $n_3$ are whole numbers in the interval $[-m,m]$$^{\ref{Foot1}}$ and such that $n_3$ can not be zero.
\begin{equation}
x_i = n_1a_x+n_2b_x+\frac{1}{2}n_3c_x \label{E16} \\
\end{equation}
\begin{equation}
y_i = n_2b_y+\frac{1}{2}n_3c_y \notag \\
\end{equation}
\begin{equation}
z_i = \frac{1}{2}n_3c_z \notag\\
\end{equation}
\end{itemize}
\item If the studied nucleus is centered in the vertex of the structure, the coordinates of the ligands are computed as follows:
\begin{itemize}
\item If the structure is simple (SC, ST, SO, SM, triclinic, trigonal or hexagonal)$^{\ref{Foot2}}$, the coordinates of the $i^{th}$ ion are computed through equations (\ref{E12}), excluding the point $(0,0,0)$. 
\item If the structure is body centered (BCC, BCT or BCO)$^{\ref{Foot2}}$, the coordinates of the $i^{th}$ ion are computed through equations (\ref{E17}), the same way as in simple structures, excluding the point $(0,0,0)$.
\begin{equation}
x_i = \frac{1}{2}n_1a_x+\frac{1}{2}n_2b_x+\frac{1}{2}n_3c_x \label{E17} \\
\end{equation}
\begin{equation}
y_i = \frac{1}{2}n_2b_y+\frac{1}{2}n_3c_y \notag \\
\end{equation}
\begin{equation}
z_i = \frac{1}{2}n_3c_z  \notag\\
\end{equation}
\item If the structure is face centered (FCC or FCO)$^{\ref{Foot2}}$, the coordinates of the $i^{th}$ ion are computed through equations (\ref{E13}), (\ref{E14}) and (\ref{E15}) for the ligands in the faces parallel to the crystallographic planes, where the numbers $n_1$, $n_2$ and $n_3$ are whole numbers in the interval $[-m,m]$$^{\ref{Foot1}}$ and such that $n_1$ y $n_2$ can not be zero. 
\item If the structure is two face centered (2FCO or 2FCM)$^{\ref{Foot2}}$, the coordinates of the $i^{th}$ ion are computed through equations (\ref{E18}), where the numbers $n_1$, $n_2$ y $n_3$ are whole numbers in the interval $[-m,m]$$^{\ref{Foot1}}$ and such that $n_1$ and $n_2$ can not be zero.  
\begin{equation}
x_i = \frac{1}{2}n_1a_x+\frac{1}{2}n_2b_x+n_3c_x \label{E18} \\
\end{equation}
\begin{equation}
y_i = \frac{1}{2}n_2b_y+n_3c_y \notag \\
\end{equation}
\begin{equation}
z_i = n_3c_z \notag\\
\end{equation}
\end{itemize}
\item The distance to the origin is computed for each ion, via the $R$ function.
\item The ligands are ordered in growing distances to the origin$^{\ref{Foot0}}$.
\item The number of ions in each layer of the nearest neighbors is counted\footnote{Indeed, the program can be used only to count the number of neighbors, their coordinates and distances to the origin for a wide range of lattices.}.
\item The valences introduced in the step b) are assigned to each layer of the nearest neighbors computed previously.
\item The components of the EFG are computed, adding$^{\ref{Foot0}}$ to each one the contribution of each ion through function $V$.
\end{enumerate}

It is important to point out that in our calculations the contribution of the Sternheimer factors to the EFG has not been taken into account, so the calculated values will be normally smaller than the experimental values. However, the purpose of these calculations is to endow a guide to discriminate between different site environments of the iron nucleus through the relative magnitudes of the calculated EFG\textcolor{NavyBlue}{\cite{Gomez}}.

\section{Conclusions}

\noindent The algorithm and the program developed here are useful as a high applicability tool in both experimental spectroscopy and in any theoretical research in solid state and crystallography, requiring this kind of computations.

The program presented in this work is extremely versatile and friendly with the final user, and only requires the structural information of the system under study. In spite of the fact that the EFG tensor and its quadrupolar splitting computations are based in a simple point charge model, disregarding the valence contribution, it is a useful tool to discriminate the different structures present in complex ionic systems.  

The program was written in Fortran 77, to assure high compatibility across different platforms, but the structure of the program and the algorithm are equally useful if the program is written in any other structured computational language, like C, Phyton, Pascal, etc., without compromising its accuracy and stability.

Of course, the program can be improved including the effect of the electronic density in the EFG tensor, but that matter may be explored in a future work. However, the computation of the shielding and antishielding factors is complicated, so their effect has to be taken into account with empirical adjustments, without changing significantly the relative magnitudes of the quadrupolar splittings.

\appendix
\section{Transformation from crystallographic to rectangular coordinates} \label{ap:coords}

\begin{figure}
\centering
\includegraphics[width=7.5cm]{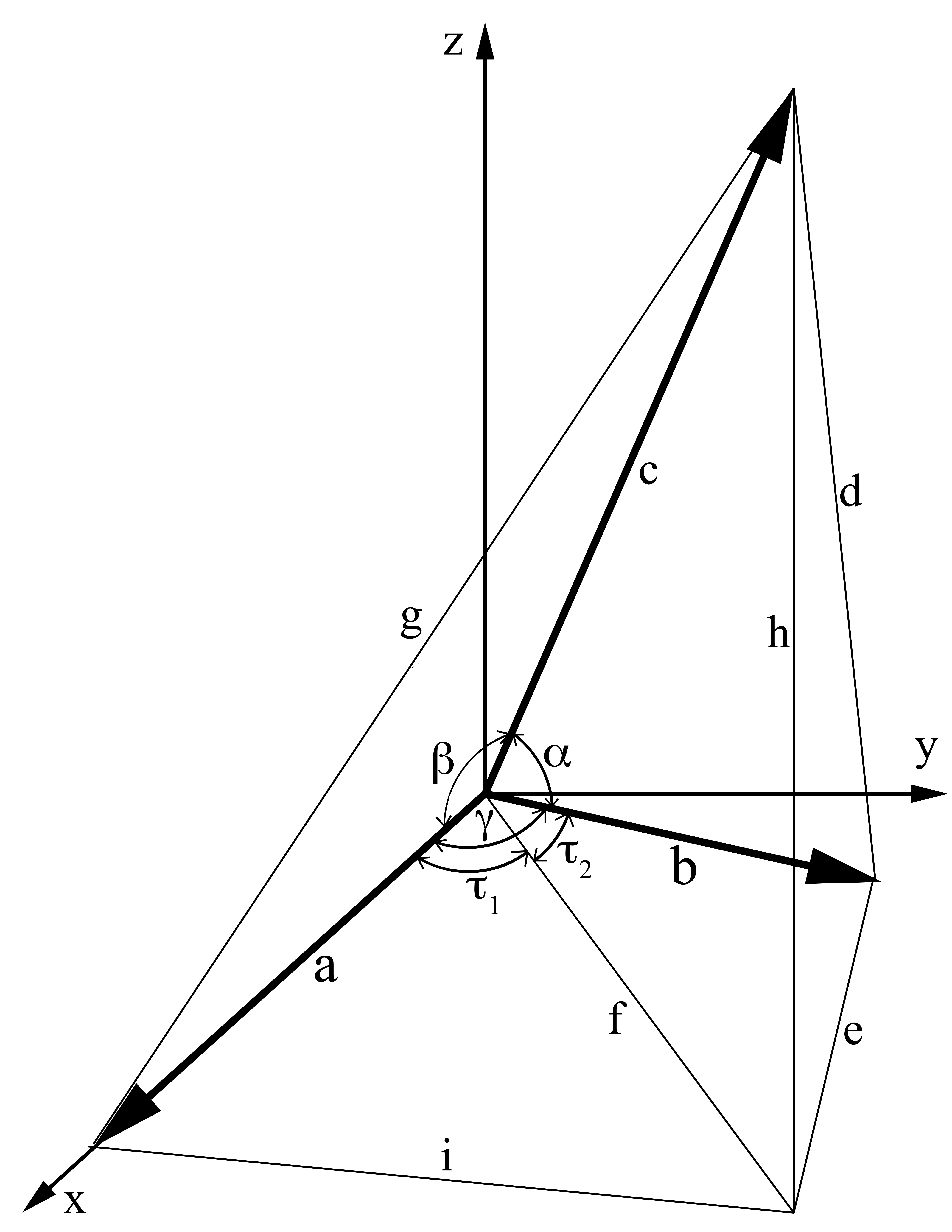} 
\caption{Geometrical scheme of the coordinate transformation}
\label{FCoords}
\end{figure}

\noindent Consider the crystallographic axes $\bar{a}$, $\bar{b}$ and $\bar{c}$, and the rectangular ones $\bar{x}$, $\bar{y}$ and $\bar{z}$, as shown in figure \ref{FCoords}.

The crystallographic axes can be expressed in the cartesian base, as:

\begin{align}
\bar{a}&=a\hat{e}_x \label{E101}\\
\bar{b}&=b\cos\gamma\hat{e}_x+b\sin\gamma\hat{e}_y \label{E102}\\
\bar{c}&=f\cos\tau_1\hat{e}_x+f\sin\tau_1\hat{e}_y+h\hat{e}_z \label{E103}\\ \notag
\end{align}

Solving the system of equations that arise from the seven triangles shown in figure \ref{FCoords}, and considering the respective constrictions, a general expression of the crystallographic axes in the rectangular base in terms of the lattice parameters can be written as

\begin{equation}
\bar{a}=a\hat{e}_x \label{E104}\\
\end{equation}
\begin{equation}
\bar{b}=b\cos\gamma\hat{e}_x+b\sin\gamma\hat{e}_y \label{E105}\\
\end{equation}
\begin{equation}
\bar{c}=c\cos\beta\hat{e}_x+c(\cos\alpha\csc\gamma-\cos\beta\cot\gamma)\hat{e}_y \label{E106}  \\
\end{equation}
\begin{equation}
+c\sin\beta\sqrt{1-(\cos\alpha\csc\beta\csc\gamma-\cot\beta\cot\gamma)^2}\hat{e}_z \notag\\ 
\end{equation}
for any non-orthonormal set of crystallographic axes.

\subsection*{Acknowledgements}
This work was partially supported by DGAPA-UNAM IN110808 grant and CONACyT, México.

\subsubsection*{Note}
The program is available for its free usage under request, if and only if the appropriate acknowledgement and credit is given to the authors.

\newpage

\tableofcontents

\listoffigures

\end{document}